\documentclass[aps,10pt,prl,twocolumn,superscriptaddress]{revtex4-1}
\usepackage{amsfonts,amssymb,amsmath,graphicx}

\newcommand{\half}{\frac{1}{2}}
\newcommand{\halfpi}{\frac{\pi}{2}}
\newcommand{\hb}{\bar{h}}
\newcommand{\rmd}{{\rm d}}

\newcommand{\Cbb}{\mathbb{C}}
\newcommand{\Rbb}{\mathbb{R}}
\newcommand{\Zbb}{\mathbb{Z}}

\newcommand{\ul}{\underline}
\newcommand{\wh}{\widehat}
\newcommand{\wt}{\widetilde}
\newcommand{\ulwh}[1]{\underline{\widehat{#1}}}

\newcommand{\myid}{{\mathchoice {\rm 1\mskip-4mu l} {\rm 1\mskip-4mu l}
{\rm 1\mskip-4.5mu l} {\rm 1\mskip-5mu l}}}

\graphicspath{{eps/}}

\begin{document}

\title{An integrable spin chain for the SL(2,R)/U(1) black hole sigma model}
\date{\today}

\author{Yacine Ikhlef}
\email{{\tt yacine.ikhlef@unige.ch}}
\affiliation{Section Math\'ematiques, Universit\'e de Gen\`eve,
  2-4 rue du Li\`evre, Case postale 64, 1211 Gen\`eve 4, Switzerland}
\author{Jesper Lykke Jacobsen}
\affiliation{Laboratoire de Physique Th\'eorique de l'Ecole Normale Sup\'erieure,
  24 rue Lhomond, 75231 Paris, France}
\affiliation{Universit\'e Pierre et Marie Curie, 4 place Jussieu, 75252 Paris, France}
\author{Hubert Saleur}
\affiliation{IPhT, CEA Saclay, 91191 Gif Sur Yvette, France}
\affiliation{USC Physics Department, Los Angeles CA 90089, USA}

\begin{abstract}
  We introduce a spin chain based on finite-dimensional spin-$1/2$
  SU(2) representations but with a non-hermitian `Hamiltonian' and
  show, using mostly analytical techniques, that it is described at
  low energies by the SL(2,R)/U(1) Euclidian black hole Conformal
  Field Theory (CFT).  This identification goes beyond the appearance
  of a non-compact spectrum: we are also able to determine the density
  of states, and show that it agrees with the formulas in
  [J. Math. Phys. 42, 2961 (2001)] and [JHEP 04, 014 (2002)], hence
  providing a direct `physical measurement' of the associated
  reflection amplitude.
\end{abstract}

\maketitle

\paragraph{\bf Introduction.}

The profound relation between quantum spin chains and quantum field
theories (QFTs) is central to modern theoretical physics.
Its simplest aspect is that the low-energy excitations of
a chain are described by a QFT in the continuum limit. Conversely, some
difficult, strongly interacting QFTs can be tackled using appropriate
(usually antiferromagnetic) spin chains, for which a large variety of
methods---including numerical---are available. The numerous success
stories using this approach include understanding the $\theta$-term in
the ${\rm O}(3)$ sigma model~\cite{HaldaneConj}, and developing
bosonization techniques from the concept of Luttinger
liquid~\cite{Giamarchi}.

The above examples all involve spin chains built with
finite-dimensional representations, and their continuum limits are
QFTs with a compact target. But many current
problems of physics are concerned with strongly curved, non-compact
targets. For instance, the CFT describing the transition between plateaux in
the two-dimensional (2D) Integer Quantum Hall Effect (IQHE) is
expected to be the low-energy limit of the non-compact 2D super sigma
model on $U(1,1|2)/[U(1|1)\times U(1|1)]$ at $\theta=\pi$. Also, the
dual of ${\cal N}=4$ SUSY gauge theory in 4D is 
closely related with a 2D sigma model on $PSU(2,2|4)/[SO(4, 1)\times SO(5)]$
\cite{Bena}. While it seems extremely hard to solve these sigma models
directly, one might hope that spin chain regularizations provide
access to some of their properties. A priori, these chains should
involve infinite-dimensional representations. In the IQHE such a chain
indeed arises in the very anisotropic limit of the
Chalker-Coddington network model, and involves alternating
highest and lowest weight representations of $psl(2|2)$
\cite{Zirnbauer}.

Unfortunately, the technical difficulties encountered in the analysis
of these infinite-dimensional spin chains are considerable. While
numerical methods based on Hilbert space truncations are
possible~\cite{Marston}, analytical approaches have stalled. Despite
much work on different aspects of the Bethe Ansatz (BA) in this
case~\cite{FaddeevKorchemsky,Belitsky}, it is not even known whether
the antiferromagnetic non-compact XXX spin chains are gapless!---nor
to what extent analyses based on coherent state representations and
analogies with the compact case~\cite{Ellis} make sense.

In this Letter we show how to construct a solvable, finite-dimensional,
antiferromagnetic spin chain whose low-energy physics is described
nevertheless by a non-compact CFT. This is obviously important
progress, since the usual BA techniques
can then be used, without insurmountable difficulties, to
understand non-compact CFTs.

We illustrate this discovery with the SL(2,$\Rbb$)/U(1) sigma model, a
gauged Wess-Zumino-Witten (WZW) model originally introduced in the context
of black holes in string theory~\cite{Witten} and later intensively studied for its
CFT features as well~\cite{Maldacena,Troost1,Troost2,FZZ,Teschner,Schomerus}.
While it is tempting to
assume that it (or the SL(2,$\Rbb$) WZW model) is the
continuum limit of a spin chain based on infinite-dimensional
representations of SL(2,$\Rbb$) or SL(2,$\Cbb$), this connection
remains presently entirely speculative. In contrast, we here show how
all the known features of the SL(2,$\Rbb$)/U(1) sigma model, including the
non-compact spectrum and the highly non-trivial density of states, are
obtainable starting from a rather modest-looking spin chain, one of
whose aspects however is {\em non-hermiticity}.

\paragraph{\bf The spin chain.}

The starting point is a $\Zbb_2$ staggered model, which was introduced
in relation with the antiferromagnetic Potts model~\cite{Saleur91,JS,IJS}.  It
is a variant of the six-vertex model but with alternating spectral
parameters $(u,u+\halfpi, \dots, u,u+\halfpi)$ and $(0,\halfpi, \dots,
0,\halfpi)$ on the horizontal and vertical lines of the square
lattice, respectively. Here, we use the Boltzmann weights:
$a=\sin(\gamma-u),b=\sin u, c=\sin \gamma$, in Baxter's
notations~\cite{Baxter}, encoded in the matrix $R_{ij}(u)$, and we
restrict to the regime $0< \gamma <\halfpi$.
The one-row transfer
matrix with periodic boundary conditions, for a system of width $2L$,
is
$$
\textstyle
t(u):={\rm Tr}_0 \ R_{0,2L}({\bar u}) R_{0,2L-1}(u)
\dots R_{02}({\bar u}) R_{01}(u) \,,
$$
where ${\bar u} := u-\pi/2$.
For simplicity, we suppose $L$ even.
In the very anisotropic limit $u \to 0$,
the equivalent quantum Hamiltonian is
\begin{equation} \label{eq:H}
  H := \half \sin 2\gamma \ \left[
    t^{-1}(0) t'(0) + t^{-1} \left(\halfpi\right) t'\left(\halfpi\right)
  \right] \,,
\end{equation}
where the prime denotes differentiation with respect to $u$.
In terms of Pauli matrices,
\begin{eqnarray}
  H &=& \sum_{j=1}^{2L} \Bigg[ 
  - \half \boldsymbol{\sigma}_j \cdot \boldsymbol{\sigma}_{j+2}
  + \sin^2 \gamma \ \sigma_j^z \sigma_{j+1}^z 
  -\half \cos 2 \gamma \ \myid
  \nonumber \\
  & & + i \sin \gamma \ (\sigma_{j-1}^z - \sigma_{j+2}^z)
  (\sigma_j^+ \sigma_{j+1}^- + \sigma_j^- \sigma_{j+1}^+) 
  \Bigg] \,. \label{eq:H-Pauli}
\end{eqnarray}
The parameter~$\gamma$ defines the quantum algebra ${\rm U}_q({\rm S\ell}_2)$ for the $R$-matrix,
through the relation $q=e^{i\gamma}$. Like in the open XXZ chain, this makes $H$ non-hermitian.
However, the low-lying states we study here all have real energies.
Among the conserved quantities of $H$, one has
\begin{itemize}
\item the $\Zbb_2$ charge $C :=  \prod_{j=1}^L c_{2j-1,2j}$,
\item the total magnetization $M:=\half \sum_{j=1}^{2L} \sigma_j^z$,
\item the `quasi-momentum'
$S:= \frac{\pi-2\gamma}{4\pi \gamma} \log \left[t(0) t^{-1} \left( \halfpi \right) \right]$, which reads
\begin{eqnarray}
  S &=& \frac{\pi-2\gamma}{4\pi \gamma} \log \left(
    \prod_{j=1}^L c_{2j,2j+1}
    \times \prod_{j=1}^L c_{2j-1,2j}
  \right) \,.
  \label{eq:S}
\end{eqnarray}
\end{itemize}
We have defined $c_{ij} = P_{ij} R_{ij}(-\pi/2)/\cos \gamma$, and $P_{ij}$ permutes the spins $i$ and $j$.
The three above operators commute with $H$ by the Yang-Baxter equations and the
six-vertex `ice rule'. Below we derive the
low-energy spectrum of~\eqref{eq:H-Pauli},
and establish a dictionary between the above conserved quantities and those
of the SL(2,$\Rbb$)/U(1) sigma model.

\paragraph{\bf Low-energy spectrum from the Bethe Ansatz.}

The model~\eqref{eq:H-Pauli} is solvable by the Bethe Ansatz, and the low-energy
states correspond to two sets of real roots $\{ \lambda_j \}_{j=1 \dots r_0}$
and $\{ \mu_j \}_{j=1 \dots r_1}$. The BA equations read~\cite{Baxter-PAF}
\begin{widetext}
\begin{eqnarray}
  \left[ \frac{\cosh(\lambda_j-i\gamma)}
    {\cosh(\lambda_j+i\gamma)} \right]^L &=&
  - \prod_{\ell=1}^{r_0}
  \frac{\sinh \half(\lambda_j-\lambda_\ell-2i\gamma)}
  {\sinh \half(\lambda_j-\lambda_\ell+2i\gamma)}
  \times \prod_{\ell=1}^{r_1}
  \frac{\cosh \half(\lambda_j-\mu_\ell-2i\gamma)}
  {\cosh \half(\lambda_j-\mu_\ell+2i\gamma)} \,, 
  \label{eq:BAE1} \\
  \left[ \frac{\cosh(\mu_j-i\gamma)}
    {\cosh(\mu_j+i\gamma)} \right]^L &=&
  - \prod_{\ell=1}^{r_0}
  \frac{\cosh \half(\mu_j-\lambda_\ell-2i\gamma)}
  {\cosh \half(\mu_j-\lambda_\ell+2i\gamma)}
  \times \prod_{\ell=1}^{r_1}
  \frac{\sinh \half(\mu_j-\mu_\ell-2i\gamma)}
  {\sinh \half(\mu_j-\mu_\ell+2i\gamma)} \,,
  \label{eq:BAE2} 
\end{eqnarray}
\end{widetext}
and the corresponding energy and momentum are
\begin{eqnarray*}
  E &=& -\sum_{j=1}^{r_0} \frac{2\sin^2 2\gamma}{\cosh 2\lambda_j + \cos 2\gamma}
  - \sum_{j=1}^{r_1} \frac{2\sin^2 2\gamma}{\cosh 2\mu_j + \cos 2\gamma} \,, \\
  p &=& -i \sum_{j=1}^{r_0} \log \frac{\cosh(\lambda_j-i\gamma)}
  {\cosh(\lambda_j+i\gamma)}
  -i \sum_{j=1}^{r_1} \log \frac{\cosh(\mu_j-i\gamma)}
  {\cosh(\mu_j+i\gamma)} \,.
\end{eqnarray*}
Bethe states are eigenstates of $M$ and $S$, with eigenvalues
\begin{equation*}
  m = L-r_0-r_1 \,, \qquad
  s = \sum_{j=1}^{r_0} s(\lambda_j)
  - \sum_{j=1}^{r_1} s(\mu_j) \,,
\end{equation*}
where
\begin{equation*}
  s(\lambda):=\frac{\pi-2\gamma}{4\pi \gamma}
  \log \frac{\cosh \lambda + \sin \gamma}{\cosh \lambda - \sin \gamma} \,.
\end{equation*}
The operator $C$ exchanges $\{ \lambda_j \}$ and $\{ \mu_j \}$.
Note that, when $\{\lambda_j\} = \{ \mu_j \}$, (\ref{eq:BAE1}--\ref{eq:BAE2}) reduce
to the BA of an XXZ model with anisotropy $\Delta=-\cos \gamma'$,
where $\gamma':=\pi-2\gamma$.

In the limit $r_0=r_1=L/2 \to \infty$, the roots that
solve~(\ref{eq:BAE1}--\ref{eq:BAE2}) form a pair of continuous
distributions $(\eta_0, \eta_1)$, defined respectively on the intervals $[-\Lambda'_0,\Lambda_0]$
and $[-\Lambda'_1,\Lambda_1]$, and subject to two coupled linear integral
equations:
\begin{equation} \label{eq:BAE-density}
  2\pi \eta_a(\lambda) + \sum_{b=0,1}
  \int_{-\Lambda'_b}^{+\Lambda_b} \rmd\mu \ \eta_b(\mu)
  K_{a-b}(\lambda-\mu) = \phi(\lambda) \,,
\end{equation}
where $a \in \{0,1\}$. It is convenient to define the kernels $K_{a-b}$ and
the source term $\phi$ through their Fourier transform, with the convention
$\wh{f}(\omega) := \int \rmd\lambda \ f(\lambda) e^{i\omega\lambda}$.
One has:
\begin{equation*}
  \wh{K}_0, \wh{K}_{\pm 1}, \wh{\phi} =
  -\frac{2\pi \sinh \gamma'\omega}{\sinh \pi\omega},
  \frac{2\pi \sinh 2\gamma \omega}{\sinh \pi\omega},
  \frac{2\pi \sinh \gamma\omega}{\sinh \frac{\pi\omega}{2}} \,.
\end{equation*}
Like in the XXZ model, the ground state (gs) corresponds to the
limit $\Lambda_a=\Lambda'_a \to \infty$ in~\eqref{eq:BAE-density}, and the solution is
simply obtained by Fourier transform
$\eta_0(\lambda) = \eta_1(\lambda) = \eta_{\rm gs}(\lambda)
:= 1/(2\gamma' \cosh \frac{\pi \lambda}{\gamma'})$.
The central charge obtained from the scaling of the ground-state
energy is $c = 2$.

The elementary excitations over the ground state (spinons)
are holes in the root distributions $\eta_0, \eta_1$.
Using standard kernel methods~\cite{YY,IKB}, we get
the dressed magnetic charge ${\cal Z}= \pi/(4\gamma)$ and
the energy and momentum of a spinon of rapidity $\lambda$:
\begin{equation*}
  \epsilon_{\rm sp}(\lambda) = -\frac{\pi \sin \gamma'}{\gamma' \cosh \frac{\pi \lambda}{\gamma'}}
  \,, \quad
  p_{\rm sp}(\lambda) = 2 {\rm Arctan} \left(\tanh \frac{\pi \lambda}{2\gamma'} \right)
  \,.
\end{equation*}
The low-energy spinons ($\lambda \to \infty$) thus have a linear dispersion,
with Fermi velocity $v_f =\frac{\pi \sin \gamma'}{\gamma'}$.
Similarly, the quasi-momentum associated to a spinon is
\begin{equation*}
  s_{\rm sp}(\lambda) = \pm \frac{\pi-2\gamma}{4\pi \gamma} \log \left[
    \cosh \frac{\pi \lambda}{\gamma'}
  \right] \,,
\end{equation*}
where the sign depends on which distribution ($\eta_0$ or $\eta_1$) the
spinon lives in. Since $s \propto (r_1-r_0)$ for large $L$ (see~\eqref{eq:mt}),
this quantity measures the difference between the two total root densities.

The main object of this Letter is the study
of the conformal spectrum $\{(h,\hb)\}$ through the finite-size behavior
of the energy gap and the momentum~\cite{Cardy}:
$$
\Delta E = E-E_{\rm gs} \simeq \frac{2\pi v_f}{L}(h+\hb) \,,
\qquad p \simeq \frac{2\pi}{L}(h-\hb) \,.
$$
In analogy with the XXZ case~\cite{Alcaraz}, we assume that
the Bethe states which converge to primaries in the
continuum limit are combinations
of a magnetic excitation (removal
of $m_0$ roots $\{\lambda_j\}$ and $m_1$ roots $\{\mu_j\}$) and an electric
excitation (global shift of the Bethe integers by an integer $e$):
such a state is then denoted $\Psi_{m_0,m_1,e}$.
It is well-known how to extract conformal weights from the BA
equations of a gapless spin chain using Wiener-Hopf (WH) analysis~\cite{YY}.
Reproducing the standard calculation~\cite{IKB} leads immediately to
the conformal weights for $\Psi_{m_0,m_1,e}$:
\begin{equation} \label{eq:h-naive}
  h+\hb = \frac{m^2}{8[1+\ulwh{J}_0(0)]}
  + 2[1+\ulwh{J}_0(0)] e^2
  + \frac{\wt{m}^2}{8[1+\ulwh{J}_1(0)]} \,,
\end{equation}
where we introduced the symmetric and anti-symmetric magnetic charges
$m:= m_0+m_1$ and $\wt{m} := m_0-m_1$, and the inverse kernels
\begin{equation}
  1+\ulwh{J}_r(\omega) := \frac{2\pi}
  {2\pi + \wh{K}_0(\omega) +(-1)^r \wh{K}_1(\omega)} \,.
\end{equation}
However, the particular feature of the $\Zbb_2$ staggered model is
that {\it the kernel $(1+\ulwh{J}_1)$ has a double pole at $\omega=0$}.
This means that, for finite $\wt{m}$, the third term in~\eqref{eq:h-naive}
vanishes identically, suggesting an infinitely degenerate
ground state in the thermodynamic limit, or, more accurately, a
continuous spectrum of exponents~\cite{IJS}. To establish this, more analysis is 
obviously needed~\cite{note1}.

We now show how to handle this problem,
emphasizing the main differences from~\cite{YY,IKB}.
Starting from~\eqref{eq:BAE-density}, our strategy consists in expanding
$m,s,\wt{m}$ and $\Delta E$ in terms of the small parameters
$\xi_a := e^{-\pi \Lambda_a / \gamma'}$ and then (as in~\cite{YY}) eliminating the $\xi_a$'s
from the equations, to get $\Delta E$ and $\wt{m}$ as functions of $(m,s,L)$.
The finite-size effects on $\Delta E$ then yield the conformal weights, whereas
the constraint $\wt{m} \in \Zbb$ determines the density of states.

We restrict for simplicity to a purely
magnetic state $\Psi_{m_0,m_1,0}$, for which
the $\eta_a$'s are even functions. 
Defining the combinations $\ul{\eta}_r:= \eta_0 + (-1)^r \eta_1$,
we may rewrite~\eqref{eq:BAE-density} as a pair of coupled WH
equations:
\begin{equation}
  \ul{\eta}_r(\lambda) + \sum_{a=0,1} (-1)^{ar}
  \int_{\Lambda_a}^\infty \eta_a(\mu) \ul{J}_r(\lambda-\mu) \rmd\mu
  = 
  2 \delta_{r,0} \ \eta_{\rm gs}(\lambda) \,.
  \label{eq:WH}
\end{equation}
We write the WH decomposition of the kernels as
$1 +\ulwh{J}_r(\omega) = [\ulwh{G}^+_r(\omega) \ulwh{G}^-_r(\omega)]^{-1}$,
where $\ulwh{G}^+_r$ (resp. $\ulwh{G}^-_r$) is analytic and
non-zero in the upper (resp.\ lower) half plane.
This is given by
\begin{equation}
  \left\{ \begin{array}{l}
      \ul{\wh{G}}_0^+(\omega) =
      \frac{\sqrt{4\gamma}
        \ \Gamma \left(1-\frac{i\omega}{2} \right)}
      {\Gamma \left(1- \frac{i\gamma\omega}{\pi} \right)
        \ \Gamma \left(\half- \frac{i \gamma' \omega}{2\pi} \right)} \,, \\
      \ul{\wh{G}}_1^+(\omega) =
      \frac{\sqrt{\frac{\gamma \gamma'}{\pi}} \ i\omega
        \ \Gamma \left(\half-\frac{i\omega}{2} \right)}
      {\Gamma \left(1- \frac{i\gamma\omega}{\pi} \right)
        \ \Gamma \left(1- \frac{i \gamma' \omega}{2\pi} \right)} \,,
    \end{array} \right.
\end{equation}
and $\ulwh{G}^-_r(\omega):=\ulwh{G}^+_r(-\omega)$.
We define the shifted densities $g_a^+(\lambda):= \eta_a(\lambda+\Lambda_a) \Theta(\lambda)$,
where $\Theta$ stands for Heaviside's step function.
The solution of~\eqref{eq:WH} can be expanded on the poles $\{ \omega_0, \omega_1, \dots \}$
of $\wh{\eta}_{\rm gs}$ in the lower half-plane, and the leading order is
\begin{equation} \label{eq:sol}
  \wh{g}_a^+(\omega) \simeq \frac{C}{\omega-\omega_0} \sum_{b=0,1}
  e^{i\omega(\Lambda_b-\Lambda_a)}
  \wh{G}_{a-b}^+(\omega) \xi_b \,,
\end{equation}
where $\wh{G}^+_{a-b}:=\half[\ulwh{G}^+_0 + (-1)^{a-b} \ulwh{G}^+_1]$, 
$\omega_0:=-{i\pi}/{\gamma'}$ and $C:=\ulwh{G}_0^-(\omega_0) {\rm Res}(\wh{\eta}_{\rm gs}, \omega_0)$.
Following~\cite{YY}, we get:
\begin{equation}
  \frac{m}{L} \simeq -\frac{2C (\xi_0+\xi_1)}{\omega_0 \ulwh{G}_0^-(0)} 
  \,, \qquad
  \frac{\Delta E}{L} \simeq 2\pi v_f
  \frac{C^2}{\omega_0^2} (\xi_0^2 + \xi_1^2)
  \,. \label{eq:m-E-xi}
\end{equation}
The derivation of $\wt{m}$ and $s$ is more involved,
due to the singularity of $\ulwh{J}_1$.
From~\eqref{eq:BAE-density}, we have
\begin{eqnarray*}
  \frac{\wt{m}}{L} &=&
  \lim_{\omega \to 0} \frac{\sum_{a=0,1}(-1)^a \left[
      e^{i\omega \Lambda_a} \wh{g}_a^+(\omega)
      + e^{-i\omega \Lambda_a} \wh{g}_a^+(-\omega)
    \right]}{\ulwh{G}_1^+(\omega) \ulwh{G}_1^-(\omega)} \,, \\
  \frac{s}{L} &=&
  -\sum_{a=0,1}  \ \frac{(-1)^a}{\pi}
  \int \rmd\omega \ \wh{s}_{\rm sp}(\omega)
  \wh{g}_a^+(\omega) e^{i\omega \Lambda_a} \,.
\end{eqnarray*}
Inserting the WH solution~\eqref{eq:sol} yields
\begin{equation}
  \wt{m} \simeq \frac{2iC(\Lambda_0 \xi_0 - \Lambda_1 \xi_1) L }
  {\omega_0 (\ulwh{G}_1^-)'(0)}
  \,, \quad
  s \simeq \frac{\gamma'^2 C (\xi_0-\xi_1) L}
  {2\pi (\ulwh{G}_1^-)'(0)} \,.
  \label{eq:mt-s-xi}
\end{equation}
Consider the regime where $m$ and $s$ are finite. Eqs.~\ref{eq:m-E-xi}--\ref{eq:mt-s-xi}
then give the scaling of $\xi_0$ and $\xi_1$:
\begin{equation*}
  \xi_0, \xi_1 \propto \frac{1}{L} \,,
  \qquad
  (\xi_0-\xi_1) \propto \frac{1}{L} \,.
\end{equation*}
Eliminating these variables from
Eqs.~\ref{eq:m-E-xi}--\ref{eq:mt-s-xi}, we obtain 
\begin{eqnarray}
  \wt{m} &\simeq& \frac{4s}{\pi} \left[
    \log \frac{L}{L_0} + B(\gamma,m, e, s)
  \right] \,,
  \label{eq:mt} \\
  \Delta E &\simeq& \frac{2\pi v_f}{L} \left(
    \frac{\gamma m^2}{2\pi}  + \frac{2\gamma s^2}{\pi-2\gamma}
  \right)
  \,. \label{eq:DeltaE}
\end{eqnarray}
In Eqs.~\ref{eq:mt}--\ref{eq:DeltaE} (derived here for $e=0$),
$L_0$ is a cut-off depending only on $\gamma$, and $B$ is a
correction term which we discuss below.
More generally, for $e \neq 0$, a similar derivation yields
\begin{equation*}
  \Delta E \simeq \frac{2\pi v_f}{L} \left(
    \frac{\gamma m^2}{2\pi} + \frac{\pi e^2}{2 \gamma}
    + \frac{2\gamma s^2}{\pi-2\gamma}
  \right) \,, \quad
  p = \frac{2\pi em}{L} \,,
\end{equation*}
and hence
\begin{equation} \label{eq:h-hb}
  h = \frac{\left( m + e k \right)^2}{4k}
  +\frac{s^2}{k-2} \,,
  \quad
  \hb = \frac{\left( m - e k \right)^2}{4k}
  +\frac{s^2}{k-2} \,,
\end{equation}
where $(m,e) \in \Zbb^2$, $s \in \Rbb$, and we have set $k:= \pi/\gamma$.
Since $s$ is real, this is a non-compact spectrum.

\paragraph{\bf Relation to the sigma model.}

Recall now that, while the SL(2,$\Rbb$)/U(1) black hole (BH) model has
central charge $c_{\rm BH}={2(k+1)}/{(k-2)}$, the identity field in
that theory is associated with a non-normalizable state. In fact,
normalizable states arise mostly from continuous representations, and
have conformal weights as in \eqref{eq:h-hb}, but with the second term
${s^2}/{(k-2)}$ replaced by a WZW-type term ${-j(j+1)}/{(k-2)}$,
with $j=-\frac{1}{2}+is$. The `bottom' of the spectrum thus occurs at
$h_0 := 1/[4(k-2)]$, leading to an effective central charge
$c=c_{\rm BH}-24h_0=2$ as in our lattice model. The spectrum
\eqref{eq:h-hb} is thus formally identical with the SL(2,$\Rbb$)/U(1)
one~\cite{Schomerus,Troost1}.

Since, in the large-$L$ limit, $s$ becomes a real parameter, the
spectrum~\eqref{eq:h-hb} is a collection of continua over the
conformal weights of a compact boson. In the SL(2,$\Rbb$)/U(1)
theory, this boson describes excitations along the compact direction
of the cigar (angular momentum of rotations around the tip),
whereas $s$ is the angular momentum along the axis of the cigar. 
We have expressed in~\eqref{eq:S} the lattice operator $S$ measuring
this angular momentum.
In finite size, since $s \simeq \pi\wt{m}/(4\log L)$,
the $s^2$ terms in~\eqref{eq:h-hb} correspond to the
magnetic charge of a boson with effective compactification
radius $R \propto \log L$.

\begin{figure}[ht]
  \begin{center}
    \includegraphics[scale=1]{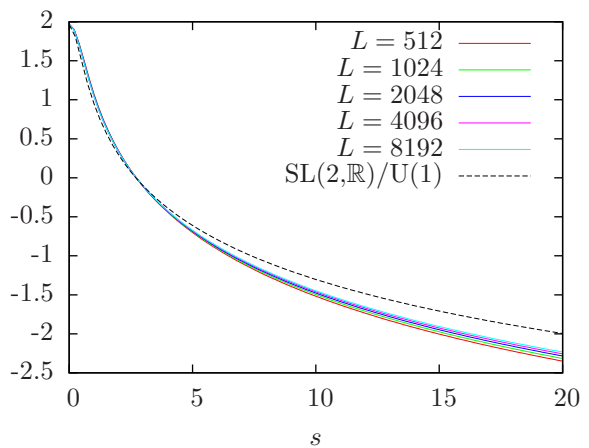}
    \caption{(Color online) Finite part $B(s)$ of the density of states~\eqref{eq:rho} for the 
      continuum over the
      ground state of the $\Zbb_2$ model at $\gamma=\pi/5$,
      compared to $B_{\rm BH}$.}
    \label{fig:B0}
  \end{center}
\end{figure}

\begin{figure}[ht]
  \begin{center}
    \includegraphics[scale=1]{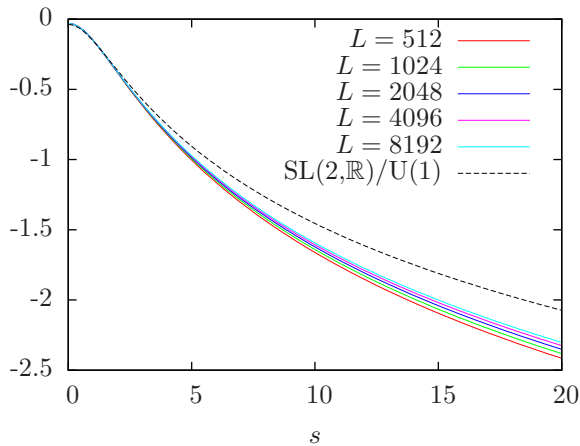}
    \caption{(Color online) Same as Fig.~\ref{fig:B0}, but for the continuum
      over the excited state $(m=2,e=0)$.}
    \label{fig:B2}
  \end{center}
\end{figure}

As in ordinary quantum mechanics, there is actually 
little dynamical information in the spectrum~\eqref{eq:h-hb} alone:
what is really needed is the {\it density of states}.
This can also be extracted from our finite-size
calculation. Denoting $q=\exp(2i\pi\tau)$ the modular parameter,
the partition function of our
model on a torus reads, in the scaling limit,
\begin{eqnarray*}
  Z &=& \frac{(q\bar{q})^{-2/24}}{|\eta(\tau)|^4}
  \sum_{e \in \Zbb, \ m+\wt{m} \in 2 \Zbb}
    q^h \ \bar{q}^{\hb} \\
  &=& \frac{(q\bar{q})^{-2/24}}{|\eta(\tau)|^4}
  \sum_{e,m \in \Zbb^2} \int_{-\infty}^{+\infty} \rmd s
  \ \rho(s) \ q^h \ \bar{q}^{\hb} \,,
\end{eqnarray*}
where $\eta$ is the Dedekind eta function, and the density of states is
\begin{equation} \label{eq:rho}
  \rho(s) = \frac{2}{\pi} \left[
    \log \frac{L}{L_0} + \partial_s(sB)
  \right] \,,
\end{equation}
where $B$ was introduced in~\eqref{eq:mt}.
The logarithmic divergence with the IR cutoff is familiar in the 
sigma model~\cite{Schomerus}, whereas the finite part of $\rho(s)$
is determined by requiring $\wt{m} \in \Zbb$ in~\eqref{eq:mt}.
Consider the purely magnetic state $\Psi_{m_0,m_1,0}$.
Our WH technique only gives access~\cite{note2}
to the function $B$ in the regime of large $s$ and $m$, where we get
\begin{equation} \label{eq:B-WH}
  B(\gamma,m,e=0,s) \sim \begin{cases}
    - \log s & \text{for $s \gg m$} \,, \\
    - \log m &  \text{for $s \ll m$} \,.
  \end{cases}
\end{equation}
We believe it will eventually be possible to obtain more complete results on $B$
by a deeper analysis of the BAE. For now, in order to interpolate
between the above limiting behaviors, we compute
$B$ numerically, by solving~(\ref{eq:BAE1}--\ref{eq:BAE2}) at
finite $L$: see results on Figs.~\ref{fig:B0}--\ref{fig:B2}.
The only adjustable parameter in these computations is $L_0$,
which can be fixed, {\it e.g.}, by imposing the value of $B$ in the ground
state $m=e=s=0$. Slow convergence with the system size is to be expected,
because higher-order corrections to Eq.~\ref{eq:mt} are of order $1/\log L$.

\begin{figure}[ht]
  \begin{center}
    \includegraphics[scale=1]{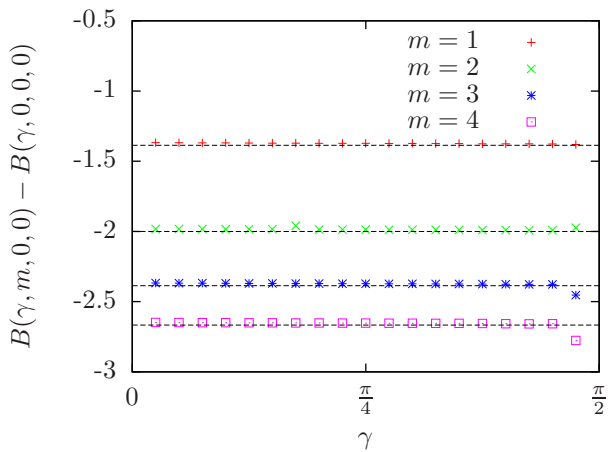}
    \caption{(Color online) The values of $B$ at $s=0$ along the interval $0<\gamma<\halfpi$.
      The data points represent extrapolated numerical values, and the dotted lines
      are the corresponding values in the SL(2,$\Rbb$)/U(1) model, taken from~\eqref{eq:B}.
    }
    \label{fig:BB-diff}
  \end{center}
\end{figure}

The finite part of the density of states $\rho(s)$ in the SL(2,$\Rbb$)/U(1) sigma model
was calculated in~\cite{Maldacena,Troost1} (see also \cite{Zamolodchikov}), and reads, in
our parametrization,
\begin{equation} \label{eq:B}
  B_{\rm BH}(s) = \frac{1}{2s}
   {\rm Im} \log\left[
  \Gamma {\textstyle\left( \frac{1-m+ek}{2}-is \right)}
  \Gamma {\textstyle\left( \frac{1-m-ek}{2}-is \right)} \right]\,.
\end{equation}
This function obeys the  asymptotic behavior~\eqref{eq:B-WH}, and
numerical agreement with the finite part in our model is good, as shown
in Figs.~\ref{fig:B0}--\ref{fig:B2}. Moreover, we have computed the
values of $(B-B_{\rm gs})$, where $B_{\rm gs}$ stands for the ground-state value of $B$,
in the limit $s \to 0$, to check that $L_0$ depends only on $\gamma$:
see Fig.~\ref{fig:BB-diff}.

To conclude, we have identified the continuum limit of our spin chain
as the SL(2,$\Rbb$)/U(1) black hole sigma model CFT~\cite{note3}, with
the level $k \in ]2,\infty[$. Obviously, this identification opens the
way to much further development. On the one hand, the spin chain can
be used to understand better the CFT structures, investigate issues
such as discrete states, conformal boundary conditions, {\it etc} -- it
will be particularly useful to study the so called DDV equations in this
context~\cite{DestrideVega}. On
the other hand, this example is not unique, since
there exists~\cite{EFS,FM} other spin chains with
finite representations and a non-compact continuum limit.
Hence, we plan in
particular to study sigma models with more complicated (super)
targets ({\it e.g.}, for the IQHE plateau theory) using this strategy.
 
Spin chains have also appeared from a different viewpoint
in the AdS/CFT conjecture~\cite{Maldacena98}. It was discovered that
many physical quantities on the gauge theory side can be related with
the spectra of quantum spin chains~\cite{Minahan}. These spectra in
turn can be studied by techniques addressing directly the low-energy
excitations~\cite{Kruczenski}, or via the BA. Recently, a
powerful machinery has been developed along those lines to obtain
results for the gauge theory at any coupling~\cite{Gromov}. It is
tempting to conjecture that spin chains such as ours might appear in
this context as the `gauge theory' side of some new interesting CFTs.

\begin{acknowledgments}
  \paragraph{\bf Acknowledgments.}
  H.S. thanks G. Korchemsky, V. Schomerus and J. Teschner for
  many discussions about non-compact spin chains and CFTs.
  The authors wish to thank C. Candu for a careful reading of
  the manuscript and helpful remarks.
  The work of J.L.J. and H.S. is supported by 
  the Agence Nationale de la Recherche (grant ANR-10-BLAN-0414).
  The work of Y.I. is supported by the European Research Council
  (grant CONFRA 228046).
\end{acknowledgments}


\begin{thebibliography}{99}
\bibitem{note1} The behavior of the BA kernels changes at $\gamma=\frac{\pi}{2}$, and the continuum
  limit of the spin chain is different: see \cite{IJS}.
\bibitem{note2} The WH method of~\cite{YY} is a finite-field method,
  strictly valid only for finite `particle densities' $m/L$ and $s/L$, and thus $m,s \propto L$. However,
  the energy is a polynomial in these densities, and hence the result~\eqref{eq:DeltaE} is
  reliable for all values of $m,s$.
\end{thebibliography}
\end{document}